\documentclass[showpacs,showkeys,preprintnumbers,amsmath,amssymb]{revtex4}

\usepackage{graphicx}

\begin{document}
\title{Degenerate three-level laser with parametric
amplifier and squeezed vacuum}
\author{Eyob Alebachew }
\email{yob_a@yahoo.com}
\author{K. Fesseha}
\affiliation{Department of Physics, Addis Ababa University, P. O.
Box 33085, Addis Ababa, Ethiopia}

\date{\today}

\begin{abstract}
Applying stochastic differential equations, we study the squeezing
and statistical properties of the cavity and output modes of a
degenerate three-level laser whose cavity contains a parametric
amplifier and coupled to a squeezed vacuum reservoir. We consider
the case in which the top and bottom levels of the three-level
cascade atoms injected into the cavity are coupled by the pump
mode emerging from the parametric amplifier. It turns out that the
presence of the squeezed vacuum reservoir and the parametric
amplifier contribute considerably to the mean photon number and
the degree of squeezing of the cavity and output modes. It appears
that almost perfect squeezing can be achieved at steady state and
at threshold for a suitable choice of parameters.
\end{abstract}



\maketitle
\section{Introduction}
The quantum properties of three-level lasers have been
investigated by several authors~\cite{1,2,3,4,5,6,7,8,9,10}. It is
found that three-level lasers can generate squeezed light under
certain conditions. The generation of squeezed light by
three-level lasers could be realized when either the atoms are
initially prepared in a coherent superposition of the top and
bottom levels ~\cite{1,5,6,7,8,9,10} or when these levels are
coupled by a strong coherent light~\cite{2,3}. More recently a
three-level laser with a parametric amplifier has been studied
when either the three-level atoms are initially prepared in a
coherent superposition of the top and bottom levels \cite{1} or
when these levels are coupled by the pump mode emerging from the
parametric amplifier \cite{2}. These studies show that the effect
of the parametric amplifier is to increase both the mean photon
number and the intracavity as well as the output mode squeezing
significantly. All previous studies have been confined to the case
for which the cavity mode of the three-level laser is coupled to a
vacuum reservoir. Moreover, apart from the squeezing spectrum of
the output mode, all attention has been paid to the calculation of
the squeezing and statistical properties of the cavity mode. Since
the output mode is accessible to measurement, it appears to be
appropriate to study the quantum properties of this mode.

In this paper we analyze the squeezing and statistical properties
of the cavity and output modes of a degenerate three-level laser
whose cavity contains a degenerate parametric amplifier and
coupled to a squeezed vacuum reservoir via a single-port mirror.
We consider the case for which the top and bottom levels of the
three-level atoms injected into the cavity are coupled by the pump
mode emerging from the parametric amplifier. We obtain stochastic
differential equations for the cavity mode variables employing the
pertinent master equation. Using the solutions of these equations,
we calculate the quadrature variance, the mean photon number, and
the power spectrum of the cavity and output modes. In addition, we
determine the squeezing spectrum of the output mode.

\section{Stochastic differential equations}
We consider a laser cavity containing a parametric amplifier and
coupled to a squeezed vacuum reservoir. Three-level atoms in a
cascade configuration are injected into the cavity at some
constant rate $r_{a}$. We denote the upper level by $|a\rangle$,
the middle level by $|b\rangle$, and the lower level by
$|c\rangle$, as shown in Fig. 1. The diploe allowed transitions
between levels $|a\rangle$ and $|b\rangle$ and between levels
$|b\rangle$ and $|c\rangle$ are resonant with the cavity mode. The
direct transitions between levels $|a\rangle$ and $|c\rangle$ are
dipole forbidden.

With the pump mode treated classically, a degenerate parametric
amplifier is describable in the interaction picture by the
Hamiltonian
\begin{equation}\label{1}
\hat H={i\varepsilon\over 2}(\hat a^{\dagger 2}-\hat a^2),
\end{equation}
in which $\varepsilon=\lambda\mu $ with $\lambda$ and $\mu$ being
respectively the coupling constant and the amplitude of the pump
mode. The equation of evolution of the density operator associated
with this Hamiltonian has the form
\begin{equation}\label{2}
{d\over dt}\hat \rho={\varepsilon\over 2}(\hat \rho\hat a^2-\hat
a^2\hat\rho+\hat a^{\dagger 2}\hat\rho-\hat\rho\hat a^{\dagger
2}).\end{equation} In addition, the Hamiltonian describing the
interaction of a three-level atom with the cavity mode and with
the pump mode emerging from the parametric amplifier has the form
\begin{align}\label{3}
\hat H =ig[\hat a^{\dagger}(|b\rangle\langle a|+ |c\rangle \langle
b|)-\hat a(|a\rangle\langle b|+|b\rangle\langle c|)] +i\frac
{\Omega} {2}(|c\rangle\langle a|-|a\rangle\langle c|),
\end{align}
where $\Omega$ is proportional to the amplitude of the pump mode
and $g$ is the atom-cavity mode coupling constant. We take the
initial state of a single three-level atom to be
\begin{equation}\label{4}
|\psi_{A}(0)\rangle=\frac{1}{\sqrt{2}}|a\rangle+\frac{1}{\sqrt{2}}|c\rangle
\end{equation}
and hence the density operator for a single atom is
\begin{equation}\label{5}
\hat\rho_{A}(0)={1\over 2}|a\rangle\langle a|
+\frac{1}{2}|a\rangle\langle c|+{1\over 2}|c\rangle\langle
a|+{1\over 2}|c\rangle\langle c|.
\end{equation}
Following the procedure developed in Ref. \cite{9}, we can show
that the master equation for the laser cavity mode coupled to a
squeezed vacuum reservoir to be
\begin{align}\label{6} \frac{d}{dt}\hat
\rho &=\mathcal{A}(2\hat a^{\dagger}\hat \rho\hat a-\hat a\hat
a^{\dagger}\hat \rho-\hat \rho\hat a\hat a^{\dagger})
+\mathcal{B}(2\hat a\hat \rho\hat a^{\dagger}-\hat a^{\dagger}\hat
a\hat \rho-\hat\rho\hat a^{\dagger}\hat
a)\notag\\&+\mathcal{C}(\hat a^{\dagger}\hat \rho\hat
a^{\dagger}+\hat a\hat\rho\hat a-\hat\rho\hat a^{\dagger 2}-\hat
a^2\hat \rho)+\mathcal{D}(\hat a^{\dagger}\hat \rho\hat
a^{\dagger}+\hat a\hat\rho\hat a-\rho\hat a^2-\hat a^{\dagger
2}\hat \rho),
\end{align} where
\begin{subequations}\label{7}
\begin{equation}\label{7a}
\mathcal{A}=\frac{\kappa N}{2}+\frac{A}
{4B}\bigg[1-\frac{3\beta}{2}+\beta^2\bigg],
\end{equation}
\begin{equation}\label{7b}
\mathcal{B}= \frac{\kappa(N+1)
}{2}+\frac{A}{4B}\bigg[1+\frac{3\beta}{2}+\beta^2\bigg],
\end{equation}
\begin{equation}\label{7c}
\mathcal{C}= -\frac{\kappa M }{2}+\frac{A}{4B}\bigg[-1+{\beta\over
2}+{\beta^2\over 2}+{\beta^3\over 2} \bigg],
\end{equation}
\begin{equation}\label{7d}
\mathcal{D}= -\frac{\kappa M }{2}+{A\over 4B}\bigg[-1-{\beta\over
2}+{\beta^2\over 2}-{\beta^3\over 2} \bigg],
\end{equation}
\begin{equation}\label{7e}
B=(1+\beta^2)(1+{\beta^2\over 4}),
\end{equation}
\begin{equation}\label{7f}
\beta=\Omega/\gamma,
\end{equation}
\begin{equation}\label{7g}
N=\sinh^{2} r,
\end{equation}
\begin{equation}\label{7h}
M=\sinh r \cosh r,
\end{equation}
\end{subequations}
\begin{equation}\label{8}
A={2g^2r_{a}\over \gamma^2}
\end{equation}
is the linear gain coefficient, $\kappa$ is the cavity damping
constant, $\gamma$ is the atomic decay rate assumed to be the same
for all the three levels, and $r$ is the squeeze parameter.
Therefore, on account of Eqs. \eqref{2} and \eqref{6} the master
equation for the cavity mode of the quantum optical system under
consideration becomes
\begin{align}\label{9}
\frac{d} {dt}\hat \rho &={\varepsilon\over 2}(\hat \rho\hat
a^2-\hat a^2\hat\rho+\hat a^{\dagger 2}\hat\rho-\hat\rho\hat
a^{\dagger 2})+ \mathcal{A}(2\hat a^{\dagger}\hat \rho\hat a-\hat
a\hat a^{\dagger}\hat \rho-\hat\rho\hat a\hat a^{\dagger})+
\mathcal{B}(2\hat a\hat \rho\hat a^{\dagger}-\hat a^{\dagger}\hat
a\hat
\rho-\hat \rho\hat a^{\dagger}\hat a)\notag\\
&+\mathcal{C}(\hat a^{\dagger}\hat \rho\hat a^{\dagger}+\hat
a\hat\rho\hat a-\hat\rho\hat a^{\dagger 2}-\hat a^2\hat \rho)+
\mathcal{D}(\hat a^{\dagger}\hat \rho\hat a^{\dagger}+\hat
a\hat\rho\hat a-\rho\hat a^2-\hat a^{\dagger 2}\hat \rho).
\end{align}

We next seek to determine, applying this master equation,
stochastic differential equations for the cavity mode variables.
To this end, applying \eqref{9} one readily finds
\begin{equation}\label{10}
{d\over dt}\langle \hat a\rangle=-(\mathcal{B}-\mathcal{A})\langle
\hat a\rangle +(\mathcal{C}-\mathcal{D}+\varepsilon)\langle \hat
a^{\dagger}\rangle,
\end{equation}
\begin{equation}\label{11}
{d\over dt}\langle \hat
a^2\rangle=-2(\mathcal{B}-\mathcal{A})\langle \hat a^2\rangle
+2(\mathcal{C}-\mathcal{D}+\varepsilon)\langle \hat
a^{\dagger}\hat a\rangle+\varepsilon-2\mathcal{D},
\end{equation}
\begin{equation}\label{12}
{d\over dt}\langle \hat a^{\dagger}\hat
a\rangle=-2(\mathcal{B}-\mathcal{A})\langle \hat a^{\dagger}\hat
a\rangle +(\mathcal{C}-\mathcal{D}+\varepsilon)(\langle \hat
a^{\dagger 2}\rangle+\langle a^2\rangle)+2\mathcal{A}
\end{equation}
and the c-number equations corresponding to these normally ordered
equations are
\begin{equation}\label{13}
{d\over dt}\langle \alpha\rangle=-(\mathcal{B}-\mathcal{A})\langle
\alpha\rangle +(\mathcal{C}-\mathcal{D}+\varepsilon)\langle
\alpha^{*}\rangle,
\end{equation}
\begin{equation}\label{14}
{d\over dt}\langle
\alpha^2\rangle=-2(\mathcal{B}-\mathcal{A})\langle \alpha^2\rangle
+2(\mathcal{C}-\mathcal{D}+\varepsilon)\langle
\alpha^{*}\alpha\rangle+\varepsilon-2\mathcal{D},
\end{equation}
\begin{equation}\label{15}
{d\over dt}\langle \alpha^{*}\alpha\rangle
=-2(\mathcal{B}-\mathcal{A})\langle \alpha^{*}\alpha\rangle
+(\mathcal{C}-\mathcal{D}+\varepsilon)(\langle \alpha^{*
2}\rangle+\langle \alpha^2\rangle)+2\mathcal{A}.
\end{equation}
Based on Eq. \eqref{13}, one can write
\begin{equation}\label{16}
{d\over dt}\alpha(t)=-(\mathcal{B}-\mathcal{A})\alpha(t)
+(\mathcal{C}-\mathcal{D}+\varepsilon)\alpha^{*}(t)+f(t),
\end{equation}
where $f(t)$ is a noise force the properties of which remain to be
determined. We note that Eq. \eqref{13} and the expectation value
of Eq. \eqref{16} will have the same form provided that
\begin{equation}\label{17}
\langle f(t)\rangle=0.
\end{equation}
It is easy to see that
\begin{equation}\label{18}
\frac{d}{dt}\langle\alpha^{2}(t)\rangle=2\langle\alpha(t)\frac{d}{dt}\alpha(t)\rangle
\end{equation}
and on substituting Eq. \eqref{16} into \eqref{18}, we get
\begin{align}\label{19}
{d\over dt}\langle \alpha^2(t)\rangle
=-2(\mathcal{B}-\mathcal{A})\langle \alpha^2(t)\rangle+2\langle
\alpha(t)f(t)\rangle+2(\mathcal{C}-\mathcal{D}+\varepsilon)\langle
\alpha^{*}(t)\alpha(t)\rangle.
\end{align}
It can also be verified in a similar manner that
\begin{align}\label{20}
{d\over dt}\langle \alpha^{*}(t)\alpha(t)\rangle
&=-2(\mathcal{B}-\mathcal{A})\langle \alpha^{*}(t)\alpha(t)\rangle
+(\mathcal{C}-\mathcal{D}+\varepsilon)(\langle \alpha^{*
2}(t)\rangle+\langle \alpha^2(t)\rangle)\notag\\&+ \langle
\alpha(t)f^{*}(t)\rangle +\langle \alpha^{*}(t) f(t)\rangle.
\end{align}
Comparison of  Eqs. \eqref{14} and \eqref{19} as well as Eqs.
\eqref{15} and \eqref{20} leads
\begin{equation}\label{21}
\langle \alpha(t)f(t)\rangle={1\over 2}(\varepsilon
-2\mathcal{D}),
\end{equation}
\begin{equation}\label{22}
\langle \alpha(t)f^{*}(t)\rangle+\langle \alpha^{*}
f(t)\rangle=2\mathcal{A}.
\end{equation}

Furthermore, one can write a formal solution of Eq. \eqref{16} as
\begin{align}\label{23}
\alpha(t)=\alpha(0)e^{-(\mathcal{B}-\mathcal{A})t}+\int_{0}^t
e^{-(\mathcal{B}-\mathcal{A})(t-t^{\prime})}[(\mathcal{C}-\mathcal{D}+\varepsilon)\alpha^{*}(t^{\prime})+f(t^{\prime})]dt^{\prime}.
\end{align}
Multiplying Eq. \eqref{23} on the left by $f(t)$ and taking the
expectation value, we have
\begin{align}\label{24}
\langle\alpha(t)f(t)\rangle&=\langle\alpha(0)f(t)\rangle
e^{-(\mathcal{B}-\mathcal{A})t}+\int_{0}^t
e^{-(\mathcal{B}-\mathcal{A})(t-t^{\prime})}[(\mathcal{C}-\mathcal{D}+\varepsilon)\langle\alpha^{*}(t^{\prime})f(t)\rangle+\langle
f(t^{\prime})f(t)\rangle]dt^{\prime}.
\end{align}
Assuming that the noise force at time $t$ does not affect the
cavity mode variables at earlier times and on taking into account
\eqref{21}, Eq. \eqref{24} can be put in the form
\begin{equation}\label{25}
\int_{0}^t e^{-(\mathcal{B}-\mathcal{A})(t-t^{\prime})}\langle
f(t^{\prime})f(t)\rangle dt^{\prime}={1\over 2}(\varepsilon
-2\mathcal{D}).\end{equation} On the basis of this result, one can
write~\cite{9}
\begin{equation}\label{26}
\langle f(t^{\prime})f(t)\rangle=(\varepsilon
-2\mathcal{D})\delta(t-t^{\prime}).
\end{equation}
It can also be established in a similar manner that
\begin{equation}\label{27}
\langle
f(t)f^{*}(t^{\prime})\rangle=2\mathcal{A}\delta(t-t^{\prime}).
\end{equation}
We note that Eqs. \eqref{26} and \eqref{27} describe the
correlation properties of the noise force $f(t)$ associated with
the normal ordering.

Now introducing a new variable defined by
\begin{equation}\label{28}
\alpha_{\pm}(t)=\alpha^{*}(t)\pm\alpha(t),
\end{equation}
we easily get with the help of (16) that
\begin{equation}\label{29}
{d\over
dt}\alpha_{\pm}(t)=-\lambda_{\mp}\alpha_{\pm}(t)+f^{*}(t)\pm
f(t),\end{equation} where
\begin{equation}\label{30}
\lambda_{\mp}=(\mathcal{B}-\mathcal{A})\mp(\mathcal{C}-\mathcal{D}+\varepsilon).
\end{equation}
The solution of Eq. \eqref{29} can be written as
\begin{equation}\label{31}
\alpha_{\pm}(t)=\alpha_{\pm}(0)e^{-\lambda_{\mp}t}+\int_{0}^{t}e^{-\lambda_{\mp}(t-t^{\prime})}(f^{*}(t^{\prime})\pm
f(t^{\prime}))dt^{\prime}.
\end{equation}
It then follows that
\begin{subequations}\label{32}
\begin{equation}\label{32a}
\alpha(t)=E_{+}(t)\alpha(0)+E_{-}(t)\alpha^{*}(0)+F(t),
\end{equation}
in which
\begin{equation}\label{32b}
E_{\pm}(t)={1\over 2}(e^{-\lambda_{-}t}\pm e^{-\lambda_{+}t}),
\end{equation}
\end{subequations}
and
\begin{subequations}\label{33}
\begin{equation}\label{33a}
F(t)=F_{+}(t)+F_{-}(t),
\end{equation}
with
\begin{equation}\label{33b}
F_{\pm}(t)={1\over
2}\int_{0}^{t}e^{-\lambda_{\mp}(t-t^{\prime})}(f^{*}(t^{\prime})\pm
f(t^{\prime}))dt^{\prime}.
\end{equation}
\end{subequations}

\section{Quadrature Variance}

In this section we seek to analyze the quadrature variance of the
cavity and output modes.

\subsection{Quadrature variance of the cavity mode}
We define the quadrature operators for the cavity mode as
\begin{equation}\label{34}
\hat a_{+}=\hat a^{\dagger}+\hat a
\end{equation}
and
\begin{equation}\label{35}
\hat a_{-}=i(\hat a^{\dagger}-\hat a).
\end{equation}
These quadrature operators satisfy the commutation relation $[\hat
a_{+},\hat a_{-}]=2i$. The variance of these quadrature operators
is expressible in terms of c-number variables associated with the
normal ordering as
\begin{equation}\label{36}
\Delta
a_{\pm}^2=1\pm\langle\alpha_{\pm}(t),\alpha_{\pm}(t)\rangle,
\end{equation}
where $\alpha_{\pm}(t)$ is given by Eq. \eqref{28}. Assuming the
cavity mode to be initially in a vacuum state and taking into
account \eqref{31} along with \eqref{17}, wee see that
\begin{equation}\label{37}
\langle\alpha_{\pm}(t)\rangle=0.
\end{equation}
Thus in view of this result, Eq. \eqref{36} reduces to
\begin{equation}\label{38}
\Delta a_{\pm}^2=1\pm\langle\alpha_{\pm}^2(t)\rangle.
\end{equation}
Furthermore, applying Eq. \eqref{29}, one easily gets
\begin{align}\label{39}
{d\over dt}\langle\alpha_{\pm}^2(t)\rangle =-2\lambda_{\mp}
\langle\alpha_{\pm}^2(t)\rangle+2\langle\alpha_{\pm}(t)f^{*}(t)\rangle
\pm 2\langle\alpha_{\pm}(t)f(t)\rangle.
\end{align}
With the aid of Eq. \eqref{31} along with \eqref{26} and
\eqref{27}, we readily obtain
\begin{equation}\label{40}
{d\over
dt}\langle\alpha_{\pm}^2(t)\rangle=-2\lambda_{\mp}\langle\alpha_{\pm}^2(t)\rangle+2(\varepsilon-2\mathcal{D}\pm
2\mathcal{A})\end{equation} and at steady state we have
\begin{equation}\label{41}
\langle\alpha_{\pm}^2(t)\rangle_{ss}={\varepsilon-2\mathcal{D}\pm
2\mathcal{A}\over \lambda_{\mp}}.
\end{equation}
Hence on account of Eqs. \eqref{41}, \eqref{30}, and
\eqref{7a}-\eqref{7h}, the quadrature variance \eqref{38} takes at
steady state the form
\begin{subequations}
\begin{equation}\label{42a}
\Delta a_{+}^2={
2\kappa(1+\beta^2)(1+\beta^2/4)e^{2r}+A(4+\beta^2) \over
2(\kappa-2\varepsilon)(1+\beta^2)(1+\beta^2/4)+A(2\beta-\beta^3)}
\end{equation}
and
\begin{equation}\label{42b}
\Delta a_{-}^2={ 2\kappa(1+\beta^2)(1+\beta^2/4)e^{-2r}+3A\beta^2
\over
2(\kappa+2\varepsilon)(1+\beta^2)(1+\beta^2/4)+A(4\beta+\beta^3)}.
\end{equation}
\end{subequations}

We next proceed to calculate the quadrature variance of the cavity
mode at threshold. We note that Eq. \eqref{29} will not have a
well-behaved solution if $\lambda_{-}< 0$. Hence we identify
$\lambda_{-}=0$ as the threshold condition. With the aid of Eq.
\eqref{30} together with \eqref{7a}-\eqref{7h}, the threshold
condition can be written as
\begin{equation}\label{43}
\varepsilon={\kappa\over 2}+{A(2\beta-\beta^3)\over
4(1+\beta^2)(1+\beta^2/4)}.
\end{equation}
Now introducing this into Eqs. \eqref{42a} and \eqref{42b}, the
quadrature variance at threshold takes the form
\begin{subequations}\label{44}
\begin{equation}\label{44a}\Delta a_{+}^2\rightarrow \infty
\end{equation}
and
\begin{equation}\label{44b}
\Delta a_{-}^2={ 2\kappa(1+\beta^2)(1+\beta^2/4)e^{-2r}+3A\beta^2
\over 4\kappa(1+\beta^2)(1+\beta^2/4)+6A\beta}.
\end{equation}
\end{subequations}

Fig. 2 indicates that the quadrature variance increases with
$\beta$ and decreases with the squeeze parameter $r$. For $r=1,
\kappa=0.8$, and $A=100$, the minimum value of the quadrature
variance is found to be $0.022$ at $\beta=0.022$. We immediately
see that the intracavity squeezing is $97.8\%$ below the vacuum
level. We also note from Fig. 3 that the presence of both the
parametric amplifier and the squeezed vacuum reservoir increases
the intracavity squeezing over and above the squeezing achievable
due to the coherently driven three-level laser \cite{2}. Moreover,
Fig. 3 clearly shows that the effect of the squeezed vacuum
reservoir is significant for relatively small values of $\beta$.

In the absence of the nonlinear crystal ($NLC$) the quantum
optical system under consideration reduces to a coherently driven
degenerate three-level laser coupled to a squeezed vacuum
reservoir. The quadrature variance for this system takes upon
setting $\varepsilon=\lambda \mu=0$ (with $\mu \neq 0$) in
\eqref{42a} and \eqref{42b} the form
\begin{subequations}\label{45}
\begin{equation}\label{45a}
\Delta a_{+}^2={
2\kappa(1+\beta^2)(1+\beta^2/4)e^{2r}+A(4+\beta^2) \over
2\kappa(1+\beta^2)(1+\beta^2/4)+A(2\beta-\beta^3)}
\end{equation}
and
\begin{equation}\label{45b}
\Delta a_{-}^2={ 2\kappa(1+\beta^2)(1+\beta^2/4)e^{-2r}+3A\beta^2
\over 2\kappa(1+\beta^2)(1+\beta^2/4)+A(4\beta+\beta^3)}.
\end{equation}
\end{subequations}

As can be seen from Fig. 4 the degree of squeezing increases with
the squeeze parameter $r$ and decreases with $\beta$. Fig. 5 also
indicates that the presence of the squeezed vacuum reservoir
increases the intracavity squeezing significantly for relatively
small values of $\beta$. For A=100, $\kappa=0.8$, and $r=1.0$, the
minimum value of the quadrature variance given by \eqref{45b}
turns out to be $0.035$ and occurs at $\beta=0.023$. We then note
from this result that the intracavity squeezing is $96.5\%$.
\subsection{Quadrature variance of the output mode}
The variance of the quadrature operators for the output mode,
defined by
\begin{equation}\label{46}
\hat a_{+,out}=\hat a^{\dagger}_{out}+\hat a_{out}
\end{equation}
and
\begin{equation}\label{47}
\hat a_{-,out}=i(\hat a^{\dagger}_{out}-\hat a_{out}),
\end{equation}
can be expressed in terms of c-number variables associated with
the normal ordering as
\begin{equation}\label{48}
\Delta a_{\pm, out}^2=1\pm\langle\alpha_{\pm,
out}(t),\alpha_{\pm,out}(t)\rangle,
\end{equation}
in which
\begin{equation}\label{49}
\alpha_{\pm,
out}(t)=\sqrt{\kappa}\alpha^{*}_{\pm}(t)-\alpha_{\pm,in}(t),
\end{equation}
and
\begin{equation}\label{50}
\alpha_{\pm,in}(t)=\frac{1}{\sqrt{\kappa}}(f_{R}^{*}(t)\pm
f_{R}(t)),
\end{equation}
with $f_{R}$ being the noise force associated with the squeezed
vacuum reservoir. This noise force has the following correlation
properties:
\begin{subequations}\label{51}
\begin{equation}\label{51a}
\langle f_{R}(t)\rangle=0,
\end{equation}
\begin{equation}\label{51b}
\langle f^{*}_{R}(t)f_{R}(t^{\prime})\rangle=\kappa
N\delta(t-t^{\prime}),
\end{equation}
\begin{equation}\label{51c}
\langle f^{*}_{R}(t)f^{*}_{R}(t^{\prime})\rangle=\langle
f_{R}(t)f_{R}(t^{\prime})\rangle=\kappa M\delta(t-t^{\prime}).
\end{equation}
\end{subequations}
In view of Eqs. \eqref{50} and \eqref{51a}, one can write
\begin{equation}\label{52}
\langle \alpha_{\pm,in}(t)\rangle=0.
\end{equation}
Thus on account of Eqs. \eqref{37} and \eqref{52}, the quadrature
variance \eqref{48} takes the form
\begin{equation}\label{53}
\Delta a_{\pm, out}^2=1\pm\langle\alpha_{\pm, out}^2(t)\rangle.
\end{equation}

Furthermore, with the aid of Eq. \eqref{49}, we find
\begin{equation}\label{54}
\langle\alpha_{\pm, out}^2(t)\rangle=\kappa
\langle\alpha_{\pm}^2(t)\rangle-2\sqrt{\kappa}\langle\alpha_{\pm}(t)\alpha_{\pm,in}(t)\rangle
+\langle\alpha_{\pm, in}^2(t)\rangle.
\end{equation}
It can be readily verified that
\begin{equation}\label{55}
\langle\alpha_{\pm}(t)\alpha_{\pm,in}(t)\rangle
=\sqrt{\kappa}(M\pm N)
\end{equation}
and
\begin{equation}\label{56}
\langle\alpha_{\pm, in}^2(t)\rangle=2(M\pm N).
\end{equation}
Upon substituting Eqs. \eqref{41}, \eqref{55}, and \eqref{56} into
\eqref{54}, we get
\begin{equation}\label{57}
\langle\alpha_{\pm,
out}^2(t)\rangle_{ss}=\kappa\bigg[{\varepsilon-2\mathcal{D}\pm
2\mathcal{A}\over \lambda_{\mp}}\bigg]+ 2(1-\kappa)(M\pm N).
\end{equation}
Therefore, with the aid of this result and along with
\eqref{7a}-\eqref{7h}, the quadrature variance takes at steady
state the form
\begin{subequations}
\begin{equation}\label{58a}
\Delta a_{+,
out}^2=1+\kappa\bigg[\frac{2(1+\beta^2)(1+\beta^2/4)[2\varepsilon+\kappa(e^{2r}-1)]
+A(4-2\beta+\beta^2+\beta^3)}{2(\kappa-2\varepsilon)(1+\beta^2)(1+\beta^2/4)+A(2\beta-\beta^3)}\bigg]+(1-\kappa)(e^{2r}-1)
\end{equation}
and
\begin{equation}\label{58b}
\Delta a_{-,
out}^2=1-\kappa\bigg[\frac{2(1+\beta^2)(1+\beta^2/4)[2\varepsilon+\kappa(1-e^{-2r})]
+A(4\beta-3\beta^2+\beta^3)}{2(\kappa+2\varepsilon)(1+\beta^2)(1+\beta^2/4)+A(4\beta+\beta^3)}\bigg]-(1-\kappa)(1-e^{-2r}).
\end{equation}
\end{subequations}
On account of Eq. \eqref{43} the quadrature variance reduces at
threshold to
\begin{subequations}
\begin{equation}\label{59a}
\Delta a_{+, out}^2\rightarrow \infty
\end{equation}
and
\begin{equation}\label{59b}
\Delta a_{-,
out}^2=1-\kappa\bigg[\frac{2(1+\beta^2)(1+\beta^2/4)(2-e^{-2r})
+A(6\beta-3\beta^2)}{4\kappa(1+\beta^2)(1+\beta^2/4)+6A\beta}\bigg]-(1-\kappa)(1-e^{-2r}).
\end{equation}
\end{subequations}
We note from Fig. 6 that the amount of squeezing of the output
mode is less than that of the cavity mode for relatively small
values of $\beta$. The minimum value of the quadrature variance of
the output mode is 0.045 at $\beta=0.022$. This shows that the
maximum squeezing of the output mode is $95.5 \%$.

We next wish to consider the case for which the nonlinear crystal
is removed from the cavity. To this end, upon setting
$\varepsilon=\lambda \mu=0$ (with $\mu \neq 0$) in Eqs.
\eqref{58a} and \eqref{58b}, the quadrature variance for this case
reduces to
\begin{subequations}
\begin{equation}\label{60a}
\Delta a_{+,
out}^2=1+\kappa\bigg[\frac{2\kappa(1+\beta^2)(1+\beta^2/4)(e^{2r}-1)
+A(4-2\beta+\beta^2+\beta^3)}{2\kappa(1+\beta^2)(1+\beta^2/4)+A(2\beta-\beta^3)}\bigg]+(1-\kappa)(e^{2r}-1)
\end{equation}
and
\begin{equation}\label{60b}
\Delta a_{-,
out}^2=1-\kappa\bigg[\frac{2\kappa(1+\beta^2)(1+\beta^2/4)(1-e^{-2r})
+A(4\beta-3\beta^2+\beta^3)}{2\kappa(1+\beta^2)(1+\beta^2/4)+A(4\beta+\beta^3)}\bigg]-(1-\kappa)(1-e^{-2r}).
\end{equation}
\end{subequations}

We notice that Eqs. \eqref{60a} and \eqref{60b} represent the
quadrature variance of the output mode of a coherently driven
three-level laser coupled to a squeezed vacuum reservoir. In Fig.
7, we plot Eqs. \eqref{45b} and \eqref{60b} versus $\beta$. These
plots also show that the degree of squeezing of the output mode is
less than that of the cavity mode. We have found that the minimum
value of the quadrature variance of the output mode to be 0.055 at
$\beta=0.023$. This indicates that the maximum squeezing of the
output mode for this case is $94.5\%$ below the vacuum level. We
observe that the degree of squeezing of the of the output mode in
the presence of the NLC is greater by $1\%$ than that without the
NLC.

\section{Squeezing spectrum}
In this section we calculate the squeezing spectrum of the output
mode. The squeezing spectrum of a single-mode light is expressible
in terms of c-number variables associated with the normal ordering
as
\begin{equation}\label{61}
S_{\pm}^{out}(\omega)=1\pm
2Re\int_{0}^{\infty}\langle\alpha_{\pm}^{out}(t),\alpha_{\pm}^{out}(t+\tau)\rangle_{ss}e^{i\omega\tau}d\tau.
\end{equation}
Thus on account of Eqs. \eqref{37} and \eqref{52}, the squeezing
spectrum can be put in the form
\begin{equation}\label{62}
S_{\pm}^{out}(\omega)=1\pm
2Re\int_{0}^{\infty}\langle\alpha_{\pm}^{out}(t)\alpha_{\pm}^{out}(t+\tau)\rangle_{ss}e^{i\omega\tau}d\tau,
\end{equation}
so that introducing Eq. \eqref{49}, we get
\begin{align}\label{63}
S_{\pm}^{out}(\omega)&=1\pm 2\kappa
Re\int_{0}^{\infty}\langle\alpha_{\pm}(t)\alpha_{\pm}(t+\tau)\rangle_{ss}e^{i\omega\tau}d\tau\mp 2\sqrt{\kappa}Re\int_{0}^{\infty}\langle\alpha_{\pm}(t)\alpha_{\pm}^{in}(t+\tau)\rangle_{ss}e^{i\omega\tau}d\tau\notag\\
&\mp
2\sqrt{\kappa}Re\int_{0}^{\infty}\langle\alpha_{\pm}^{in}(t)\alpha_{\pm}(t+\tau)\rangle_{ss}e^{i\omega\tau}d\tau
\pm
2Re\int_{0}^{\infty}\langle\alpha_{\pm}^{in}(t)\alpha_{\pm}^{in}(t+\tau)\rangle_{ss}e^{i\omega\tau}d\tau.
\end{align}
Furthermore, the solution of Eq. \eqref{29} can also be written as
\begin{align}\label{64}
\alpha_{\pm}(t+\tau)= \alpha_{\pm}(t)
e^{-\lambda_{\mp}\tau}+\int_{0}^{\tau}e^{-\lambda_{\mp}(\tau-\tau^{\prime})}(f^{*}(t+\tau^{\prime})
\pm f(t+\tau^{\prime}))d\tau^{\prime}.
\end{align}
Now applying this equation along with the quantum regression
theorem, one can establish that
\begin{equation}\label{65}
\langle\alpha_{\pm}(t)\alpha_{\pm}(t+\tau)\rangle_{ss}=\langle\alpha_{\pm}^2(t)\rangle_{ss}
e^{-\lambda_{\mp}\tau},
\end{equation}
\begin{equation}\label{66}
\langle\alpha_{\pm}(t)\alpha_{\pm}^{in}(t+\tau)\rangle_{ss}=0,
\end{equation}
\begin{equation}\label{67}
\langle\alpha_{\pm}^{in}(t)\alpha_{\pm}(t+\tau)\rangle_{ss}=2\sqrt{\kappa}
(M\pm N)e^{-\lambda_{\mp}\tau},
\end{equation}
\begin{equation}\label{68}
\langle\alpha_{\pm}^{in}(t)\alpha_{\pm}^{in}(t+\tau)\rangle_{ss}=2(M\pm
N)\delta(\tau).
\end{equation}
In view of these results, the squeezing spectrum takes the form
\begin{align}\label{69}
S_{\pm}^{out}(\omega)&=1\pm 2\kappa
\langle\alpha_{\pm}^2(t)\rangle_{ss}
Re\int_{0}^{\infty}e^{-(\lambda_{\mp}-i\omega\tau)}d\tau
\mp4\kappa(M\pm N)Re
\int_{0}^{\infty}e^{-(\lambda_{\mp}-i\omega\tau)}d\tau
\notag\\&\pm 4(M\pm N)
Re\int_{0}^{\infty}e^{-i\omega\tau}\delta(\tau)d\tau.
\end{align}
Thus upon performing the integration and using \eqref{41}, we
easily find
\begin{equation}\label{70}
S_{\pm}^{out}(\omega)=1\pm
\frac{2\kappa(\varepsilon-2\mathcal{D}\pm
2\mathcal{A})-4\kappa(M\pm
N)\lambda_{\mp}}{\lambda_{\mp}^2+\omega^2} \pm 2(M\pm N).
\end{equation}
Employing Eqs. \eqref{30} and \eqref{7a}-\eqref{7g}, the squeezing
spectrum can be written as
\begin{subequations}\label{71}
\begin{equation}\label{71a}
S_{+}^{out}(\omega)=e^{2r}\bigg[1+
\frac{2\kappa[\varepsilon+\frac{A(-2\beta+\beta^3)}{(1+\beta^2)(4+\beta^2)}]+\frac{\kappa
A(4+\beta^2)e^{-2r}}{2(1+\beta^2)(1+\beta^2/4)}}
{[\frac{\kappa}{2}-\varepsilon+\frac{A(2\beta-\beta^3)}{(1+\beta^2)(4+\beta^2)})]^2+\omega^2}\bigg]
\end{equation}
and
\begin{equation}\label{71b}
S_{-}^{out}(\omega)=e^{-2r}\bigg[1-
\frac{2\kappa[\varepsilon+\frac{A(4\beta+\beta^3)}{(1+\beta^2)(4+\beta^2)}]-\frac{3\kappa
A\beta^2e^{2r}}{2(1+\beta^2)(1+\beta^2/4)}}
{[\frac{\kappa}{2}+\varepsilon+\frac{A(4\beta+\beta^3)}{(1+\beta^2)(4+\beta^2)})]^2+\omega^2}\bigg],
\end{equation}
\end{subequations}
so that on account of \eqref{43} the squeezing spectrum reduces at
threshold to
\begin{subequations}\label{72}
\begin{equation}\label{72a}
S_{+}^{out}(\omega)=e^{2r}\bigg[1+ \frac{\kappa^2+\frac{\kappa
A(4+\beta^2)e^{-2r}}{2(1+\beta^2)(1+\beta^2/4)}} {\omega^2}\bigg]
\end{equation}
and
\begin{equation}\label{72b}
S_{-}^{out}(\omega)=e^{-2r}\bigg[1- \frac{\kappa^2+\frac{3\kappa
A\beta}{(1+\beta^2)(1+\beta^2/4)}-\frac{3\kappa
A\beta^2e^{2r}}{2(1+\beta^2)(1+\beta^2/4)}}
{[\kappa+\frac{3A\beta}{(1+\beta^2)(2+\beta^2/2)})]^2+\omega^2}\bigg].
\end{equation}
\end{subequations}
In Fig. 8, we plot the squeezing spectrum [Eq. \eqref{72b}] versus
$\omega$ and $\beta$. This plot indicates that perfect squeezing
is attainable for $\omega=0$, $\beta=0$ and for any values of $A$,
$r$, and $\kappa$. To see the effect of the squeezed vacuum
reservoir on the squeezing spectrum, we plot Eq. \eqref{72b}
versus $\beta$ for $\omega=0$, $\kappa=0.8$, $A=100$, and for
different values of the squeeze parameter $r$. It is easy to see
from Fig. 9 that the presence of the squeezed vacuum reservoir
increases the degree of squeezing significantly for $\beta \neq
0$.

It is interesting to consider once more the case for which the
nonlinear crystal is removed from the laser cavity. Thus upon
setting $\varepsilon=\lambda \mu=0$ (with $\mu \neq 0$) in Eqs.
\eqref{71a} and \eqref{71b}, we obtain the squeezing spectrum to
be
\begin{subequations}\label{73}
\begin{equation}\label{73a}
S_{+}^{out}(\omega)=e^{2r}\bigg[1+
\frac{2\kappa[\frac{A(-2\beta+\beta^3)}{(1+\beta^2)(4+\beta^2)}]+\frac{\kappa
A(4+\beta^2)e^{-2r}}{2(1+\beta^2)(1+\beta^2/4)}}
{[\frac{\kappa}{2}+\frac{A(2\beta-\beta^3)}{(1+\beta^2)(4+\beta^2)})]^2+\omega^2}\bigg]
\end{equation}
and
\begin{equation}\label{73b}
S_{-}^{out}(\omega)=e^{-2r}\bigg[1-
\frac{2\kappa[\frac{A(4\beta+\beta^3)}{(1+\beta^2)(4+\beta^2)}]-\frac{3\kappa
A\beta^2e^{2r}}{2(1+\beta^2)(1+\beta^2/4)}}
{[\frac{\kappa}{2}+\frac{A(4\beta+\beta^3)}{(1+\beta^2)(4+\beta^2)})]^2+\omega^2}\bigg].
\end{equation}
\end{subequations}
Expressions \eqref{73a} and \eqref{73b} represent the squeezing
spectrum of a coherently driven three-level laser coupled to a
squeezed vacuum reservoir. In Fig. 10, we plot Eq. \eqref{73b}
versus $\beta$ for $\omega=0$, $\kappa=0.8$, $A=100$, and for
different values of the squeeze parameter $r$. These plots show
that the degree of squeezing of the output mode increases with the
squeeze parameter $r$ and almost perfect squeezing occurs for
small values $\beta$.

\section{Photon statistics}
We now proceed to calculate the mean photon number of the cavity
and output modes. The mean photon number of the cavity mode can be
written as
\begin{equation}\label{74}
\bar n=\langle\alpha^{*}(t)\alpha(t)\rangle.
\end{equation}
Employing \eqref{32a} and its complex conjugate and assuming that
the cavity mode is initially in a vacuum state, we get
\begin{equation}\label{75}
\langle\alpha^{*}(t)\alpha(t)\rangle=\langle F^{*}(t)F(t)\rangle
\end{equation}
It can be readily established using Eqs. \eqref{33a} and
\eqref{33b} along with \eqref{26} and \eqref{27} that
\begin{align}\label{76}
\langle F^{*}(t)F(t)\rangle
={2\mathcal{A}-2\mathcal{D}+\varepsilon\over
4\lambda_{-}}(1-e^{-2\lambda_{-}t
})+{2\mathcal{A}+2\mathcal{D}-\varepsilon\over
4\lambda_{+}}(1-e^{-2\lambda_{+}t }).
\end{align}
On account of Eqs. \eqref{7a}-\eqref{7h}, \eqref{30}, \eqref{75}
and \eqref{76}, the mean photon number turns out to be
\begin{align}\label{77}
\bar n
=&{[2\varepsilon+\kappa(e^{2r}-1)](1+\beta^2)(1+\beta^2/4)+A(2-\beta+
\beta^2/2+\beta^3/2)\over
4(1+\beta^2)(1+\beta^2/4)(\kappa-2\varepsilon)+2A(2\beta-\beta^3)}
(1-e^{-2\lambda_{-}t})\notag\\
-& {[2\varepsilon+\kappa(e^{-2r}-1)](1+\beta^2)(1+\beta^2/4)
+A(2\beta-3\beta^2/2+\beta^3/2)\over
4(1+\beta^2)(1+\beta^2/4)(\kappa+2\varepsilon)+2A(4\beta+\beta^3)}
(1-e^{-2\lambda_{+}t }).
\end{align}

Fig. 11 shows that the presence of both the parametric amplifier
and the squeezed vacuum reservoir increase the mean photon number
significantly for relatively small values of $\beta$. On the other
hand, it can be readily established that the mean photon number of
the output mode is expressible as \cite{9}
\begin{equation}\label{78}
\bar n_{out}=\kappa \bar n +(1-\kappa)N,
\end{equation}
where $\bar n$ is given by Eq. \eqref{77}. Fig. 12 shows that the
mean photon number of the cavity mode is greater than that of the
output mode.

\section{Power spectrum}
We finally calculate the power spectrum of the cavity and output
modes.
\subsection{Power spectrum of the cavity mode}
The power spectrum of a single-mode light is expressible in terms
of c-number variables associated with the normal ordering as
\begin{equation}\label{79}
S(\omega)=2Re\int_{0}^{\infty}\langle
\alpha^{*}(t)\alpha(t+\tau)\rangle_{ss}e^{i\omega\tau}d\tau.
\end{equation}
The solution of Eq. \eqref{29} can also be written as
\begin{subequations}\label{80}
\begin{equation}\label{80a}
\alpha(t+\tau)=E_{+}(\tau)\alpha(t)+E_{-}(\tau)\alpha^{*}(t)+F(t+\tau),
\end{equation}
in which
\begin{equation}\label{80b}
E_{\pm}(\tau)={1\over 2}(e^{-\lambda_{-}\tau}\pm
e^{-\lambda_{+}\tau})
\end{equation}
\end{subequations}
and
\begin{subequations}\label{81}
\begin{equation}\label{81a}
F(t+\tau)=F_{+}(t+\tau)+F_{-}(t+\tau),
\end{equation}
with
\begin{equation}\label{81b}
F_{\pm}(t+\tau)={1\over
2}\int_{0}^{t}e^{-\lambda_{\mp}(\tau-\tau^{\prime})}(f^{*}(t+\tau^{\prime})\pm
f(t+\tau^{\prime}))d\tau^{\prime}.
\end{equation}
\end{subequations}
Upon multiplying Eq. \eqref{80a} by $\alpha^{*}(t)$ and taking the
expectation value of the resulting expression, we have
\begin{equation}\label{82}
\langle\alpha^{*}(t)\alpha(t+\tau)\rangle_{ss}=E_{+}\langle\alpha^{*}(t)\alpha(t)
\rangle_{ss}+E_{-}\langle\alpha^{* 2}(t)\rangle_{ss}.
\end{equation}
It can be readily established using Eqs. \eqref{32a}, \eqref{32b},
\eqref{33a}, and \eqref{33b} together with the properties of the
noise force $f(t)$ that
\begin{align}\label{83}
\langle \alpha^2\rangle _{ss}=\langle \alpha^{*
2}\rangle_{ss}={2\mathcal{A}-2\mathcal{D}+\varepsilon\over
4\lambda_{-}}-{2\mathcal{A}+2\mathcal{D}-\varepsilon\over
4\lambda_{+}}.
\end{align}
In view of Eqs. \eqref{76}, \eqref{80b}, and \eqref{83},
expression \eqref{82} takes the form
\begin{align}\label{84}
\langle\alpha^{*}(t)\alpha(t+\tau)\rangle_{ss}=
\frac{2(\mathcal{A}-\mathcal{D})+\varepsilon}
{4\lambda_{-}}e^{-\lambda_{-}\tau}
+\frac{2(\mathcal{A}+\mathcal{D})-\varepsilon}{4\lambda_{+}}e^{-\lambda_{+}\tau}.
\end{align}
Applying this result in \eqref{79} and performing the integration,
the power spectrum of the cavity mode is found to be
\begin{equation}\label{85}
S(\omega)=\frac{\mathcal{A}-\mathcal{D}+\frac{\varepsilon}{2}}{\lambda_{-}^2+\omega^2}+
\frac{\mathcal{A}+\mathcal{D}-\frac{\varepsilon}{2}}{\lambda_{+}^2+\omega^2}.
\end{equation}
Employing Eqs. \eqref{7a}-\eqref{7h} and \eqref{30}, the power
spectrum can be written in the form
\begin{align}\label{86}
S(\omega)=\frac{\frac{\kappa}{4}(e^{2r}-1)+\frac{\varepsilon}{2}+
\frac{A(4-2\beta+\beta^2+\beta^3)} {(1+\beta^2)(8+2\beta^2)}}
{[\frac{\kappa}{2}-\varepsilon+\frac{A(2\beta-\beta^3)}{(1+\beta^2)
(4+\beta^2)}]^2+\omega^2}
+\frac{\frac{\kappa}{4}(e^{-2r}-1)-\frac{\varepsilon}{2}
+\frac{A(-4\beta+3\beta^2-\beta^3)} {(1+\beta^2)(8+2\beta^2)}}
{[\frac{\kappa}{2}+\varepsilon+\frac{A(4\beta+\beta^3)}
{(1+\beta^2)(4+\beta^2)}]^2+\omega^2}.
\end{align}
We see that the power spectrum is a sum of two Lorentzians
centered at zero frequency, with a halfwidth of
$\frac{\kappa}{2}-\varepsilon+\frac{A(2\beta-\beta^3)}{(1+\beta^2)
(4+\beta^2)}$ and
$\frac{\kappa}{2}+\varepsilon+\frac{A(4\beta+\beta^3)}{(1+\beta^2)
(4+\beta^2)}$. We realize that halfwidth of the two Lorentzians
does not depend on the squeeze parameter. However, the halfwidth
depends on $\varepsilon$ which represents the parametric
amplifier. In Fig. 13 we plot the power spectrum of the cavity
mode given by Eq. \eqref{86} versus $\omega$ for different values
of $\varepsilon$. These plots show that the halfwidth decreases
with $\varepsilon$. When the value of $\varepsilon$ increases from
0.2 to 0.3 the halfwidth decreases from 0.80 to 0.75.
\subsection{Power spectrum of the output mode}
The power spectrum of the output mode is expressible in terms of
c-number variables associated with the normal ordering as
\begin{equation}\label{87}
S^{out}(\omega)=2Re
\int_{0}^{\infty}\langle\alpha^{*}_{out}(t)\alpha_{out}(t+\tau)\rangle_{ss}e^{i\omega\tau}d\tau,
\end{equation}
where
\begin{equation}\label{88}
\alpha_{out}(t)=\sqrt{\kappa}\alpha(t)-\alpha_{in}(t).
\end{equation}
Now with the help of Eq. \eqref{88}, we can write
\begin{align}\label{89}
\langle\alpha^{*}_{out}(t)\alpha_{out}(t+\tau)\rangle_{ss}&=
\kappa\langle\alpha^{*}(t)\alpha(t+\tau)\rangle_{ss}-\sqrt{\kappa}\langle\alpha^{*}(t)\alpha_{in}(t+\tau)\rangle_{ss}
\notag\\&-\sqrt{\kappa}\langle\alpha^{*}_{in}(t)\alpha(t+\tau)\rangle_{ss}
+\langle\alpha^{*}_{in}(t)\alpha_{in}(t+\tau)\rangle_{ss}.
\end{align}
It can be readily verified that
\begin{equation}\label{90}
\sqrt{\kappa}\langle\alpha^{*}(t)\alpha_{in}(t+\tau)\rangle_{ss}=0,
\end{equation}
\begin{equation}\label{91}
\sqrt{\kappa}\langle\alpha^{*}_{in}(t)\alpha(t+\tau)\rangle_{ss}=
\frac{\kappa}{2}[(M+N)e^{-\lambda_{-}\tau}+(M-N)e^{-\lambda_{+}\tau}],
\end{equation}
\begin{equation}\label{92}
\langle\alpha^{*}_{in}(t)\alpha_{in}(t+\tau)\rangle_{ss}= N
\delta(\tau).
\end{equation}
Thus on account of Eqs. \eqref{84}, \eqref{90}, \eqref{91}, and
\eqref{92}, Eq. \eqref{89} takes the form
\begin{align}\label{93}
\langle\alpha^{*}_{out}(t)\alpha_{out}(t+\tau)\rangle_{ss}&=\kappa\bigg[
\frac{2(\mathcal{A}-\mathcal{D})+\varepsilon}{4\lambda_{-}}-\frac{M+N}{2}\bigg]e^{-\lambda_{-}\tau}
\notag\\&+\kappa\bigg[\frac{2(\mathcal{A}+\mathcal{D})-\varepsilon}{4\lambda_{+}}-\frac{M-N}{2}\bigg]e^{-\lambda_{+}\tau}
+N\delta(\tau).
\end{align}
On substituting this result into \eqref{87} and carrying out the
integration, we get
\begin{equation}\label{94}
S^{out}(\omega)=
\frac{\mathcal{A}-\mathcal{D}+\frac{\varepsilon}{2}-\lambda_{-}(M+N)\big]}{\lambda_{-}^2+\omega^2}
+\frac{\kappa\big[\mathcal{A}+\mathcal{D}-\frac{\varepsilon}{2}-\lambda_{+}(M-N)\big]}{\lambda_{+}^2+\omega^2}
+N.
\end{equation}
With the aid of Eqs. \eqref{30} and \eqref{7a}-\eqref{7h}, we find
the power spectrum of the output mode to be
\begin{align}\label{95}
S^{out}(\omega)&=\frac{\kappa\big[(\frac{\varepsilon}{2}-\frac{A(2\beta-\beta^3)}{(1+\beta^2)(8+2\beta^2)})e^{2r}+
\frac{A(4+\beta^2)}{(1+\beta^2)(8+2\beta^2)}\big]}
{[\frac{\kappa}{2}-\varepsilon+\frac{A(2\beta-\beta^3)}{(1+\beta^2)
(4+\beta^2)}]^2+\omega^2}
\notag\\&+\frac{\kappa\big[-(\frac{\varepsilon}{2}+\frac{A(4\beta+\beta^3)}{(1+\beta^2)(8+2\beta^2)})e^{-2r}
+\frac{3A\beta^2}{(1+\beta^2)(8+2\beta^2)}\big]}
{[\frac{\kappa}{2}+\varepsilon+\frac{A(4\beta+\beta^3)}
{(1+\beta^2)(4+\beta^2)}]^2+\omega^2}+\sinh^2{r}.
\end{align}
Expression \eqref{95} indicates that the power spectrum consists
of two Lorentzians and a flat spectrum. Equations. \eqref{86} and
\eqref{95} show that the width of the Lorentzians in the power
spectrum of the cavity and the output modes are the same.

\section{Conclusion}
We have considered a degenerate three-level laser containing a
parametric amplifier and coupled to a squeezed vacuum reservoir.
Applying the pertinent master equation, we have obtained
stochastic differential equations associated with the normal
ordering. Using the solutions of these equations, we have
calculated the quadrature variance, the mean photon number, and
the power spectrum of the cavity and output modes of the system
under consideration. We have also determined the squeezing
spectrum of the output mode.

We have seen that the effect of the squeezed vacuum reservoir and
the parametric amplifier is to increase the mean photon number and
the degree of squeezing of the cavity and output modes. We have
also found that the squeezed vacuum reservoir increases the degree
of squeezing significantly over and above the squeezing attainable
from a degenerate three-level laser with a parametric amplifier
\cite{2}. It turns out that the degree of squeezing of the cavity
mode is greater than that of the output mode for certain values of
$\beta$. On the other hand, the plots of the squeezing spectrum at
threshold show that there is perfect squeezing of the output mode
for $\beta=\omega=0$ and for any values of A, $\kappa$, and the
squeeze parameter $r$. We have also found that the mean photon
number of the cavity mode is greater that of the output mode.
Moreover, we have seen that the presence of the parametric
amplifier leads to a decrease in the width of the power spectrum
while the squeezed vacuum reservoir has no effect on the width.

\begin{acknowledgments}
EA would like to thank the Abdus Salam ICTP for the financial
support under the affiliated center (AC-14) at the Department of
Physics of the Addis Ababa University. \end{acknowledgments}

\begin{figure}[htbp]
\centering
\includegraphics [width=8.3cm]{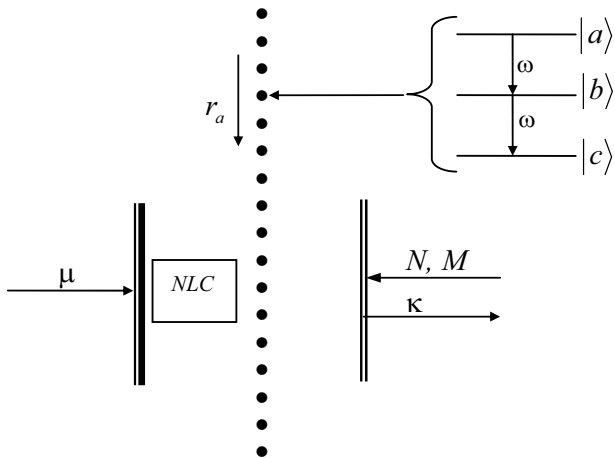}
\caption{A degenerate three-level laser with a parametric
amplifier and a squeezed vacuum reservoir.}
\end{figure}

\begin{figure}[htbp]
\centering
\includegraphics [width=8.3cm]{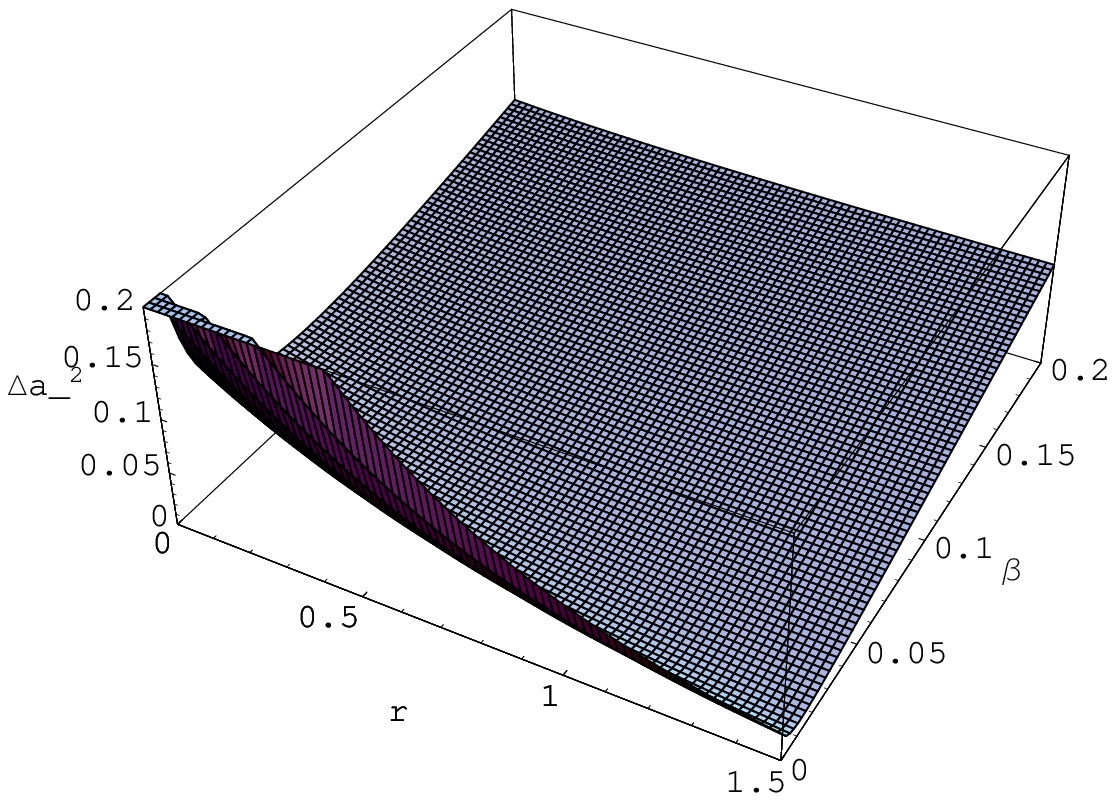}
\caption{A plot of the quadrature variance [Eq. \eqref{44b}]
versus $\beta$ and $r$ for $\kappa=0.8$ and $A=100$.}
\end{figure}

\begin{figure}[htbp]
\centering
\includegraphics [width=8.3cm]{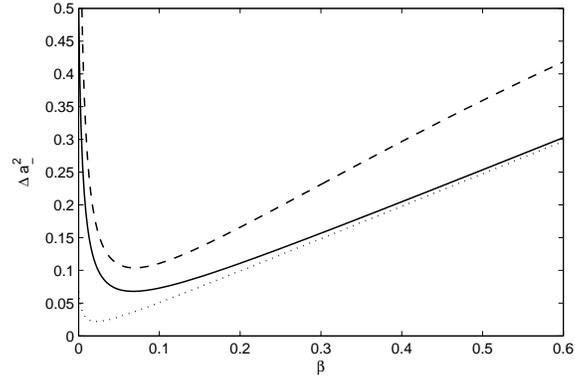}
\caption{Plots of the quadrature variance versus $\beta$ for
$\kappa=0.8$ and $A=100$, for $r=0$ and $\varepsilon=0$ [Eq.
\eqref{45b}] (dashed curve), for $r=0$ in the presence of the
parametric amplifier at threshold [Eq. \eqref{44b}] (solid curve),
and for $r=1.0$ and in the presence of the parametric amplifier at
threshold [Eq. \eqref{44b}] (dotted curve).}
\end{figure}

\begin{figure}[htbp]
\centering
\includegraphics [width=8.3cm]{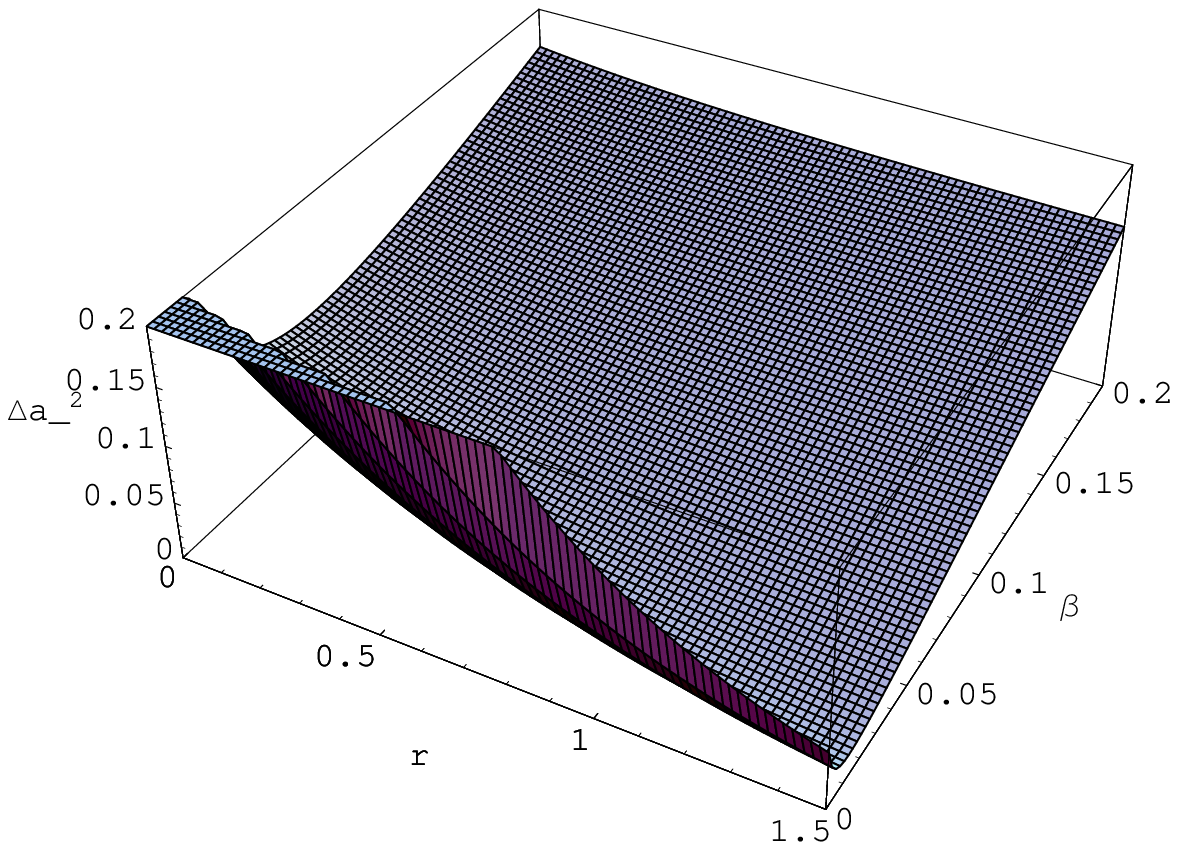}
\caption{A plot of the quadrature variance [Eq. \eqref{45b}]
versus $r$ and $\beta$ for $\kappa=0.8$ and A=100.}
\end{figure}

\begin{figure}[htbp]
\centering
\includegraphics [width=8.3cm]{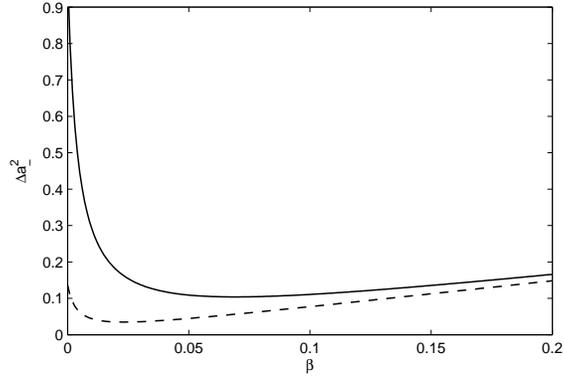}
\caption{Plots of the quadrature variance [Eq. \eqref{45b}] versus
$\beta$ for $\kappa=0.8$ and $A=100$ in the absence of the
squeezed vacuum reservoir (solid curve) and in the presence of the
squeezed vacuum reservoir with $r=1.0$ (dashed curve).}
\end{figure}

\begin{figure}[htbp]
\centering
\includegraphics [width=8.3cm]{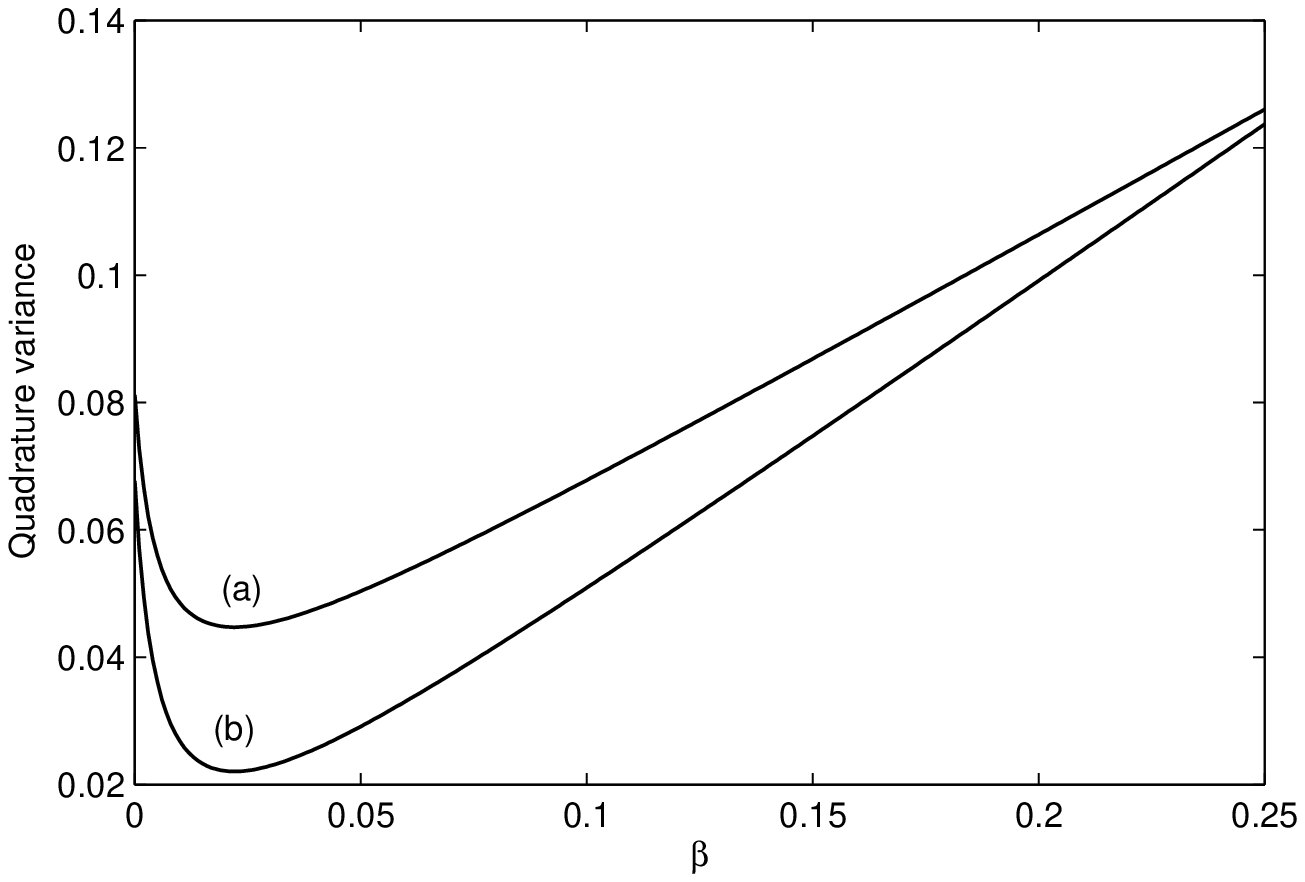}
\caption{(a) A plot of the quadrature variance of the output mode
[Eq. \eqref{59b}] versus $\beta$ for $\kappa=0.8$, $A=100$, and
$r=1$. (b) A plot of  quadrature variance of the cavity mode [Eq.
\eqref{44b}] versus $\beta$ for $\kappa=0.8$, A=100, and $r=1$.}
\end{figure}

\begin{figure}[htbp]
\centering
\includegraphics [width=8.3cm]{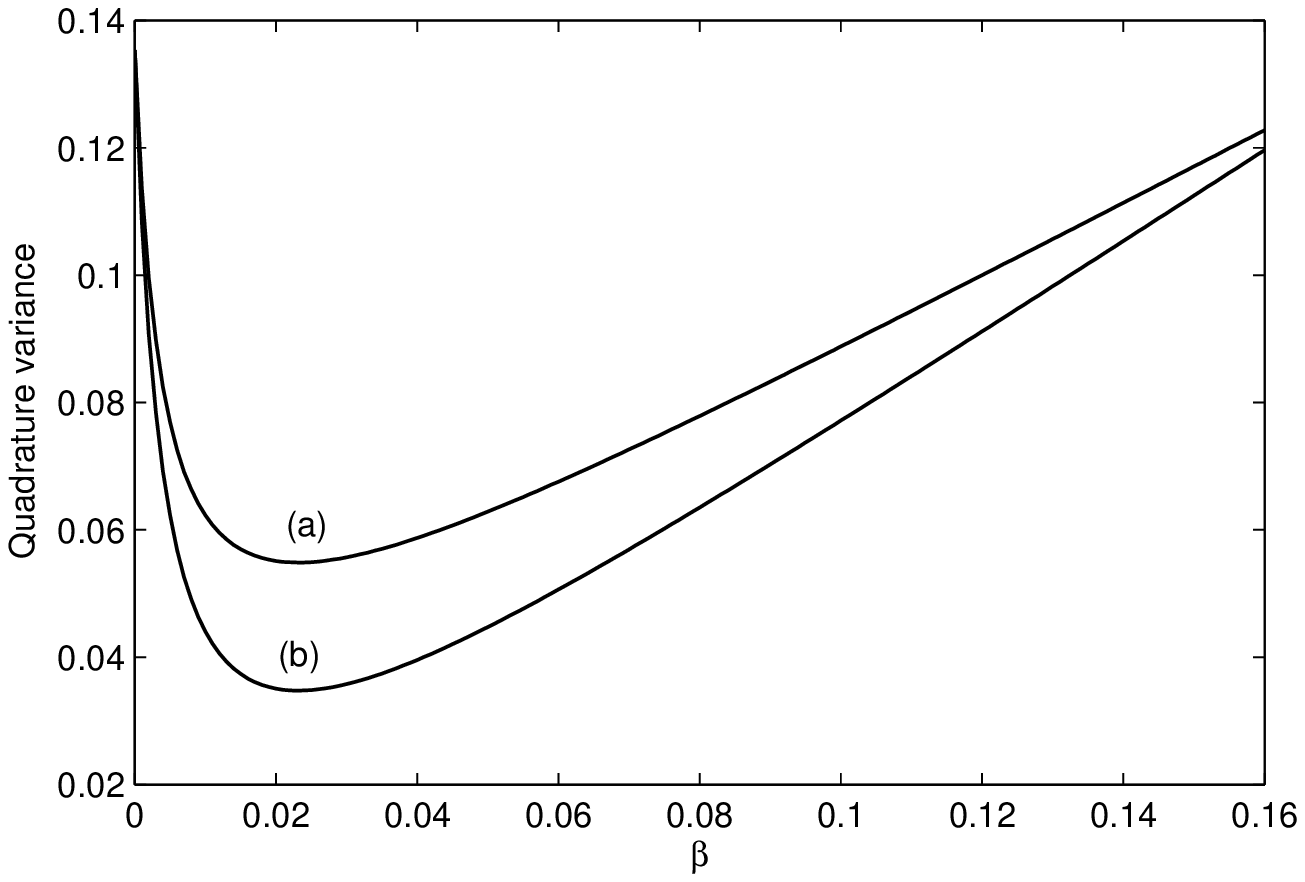}
\caption{(a) A plot of the quadrature variance of the output mode
[Eq. \eqref{60b}] versus $\beta$ for $\kappa=0.8$, $A=100$, and
$r=1$. (b) A plot of quadrature variance of the cavity mode [Eq.
\eqref{45b}] versus $\beta$ for $\kappa=0.8$, A=100, and $r=1$.}
\end{figure}
\begin{figure}[htbp]
\centering
\includegraphics [width=8.3cm]{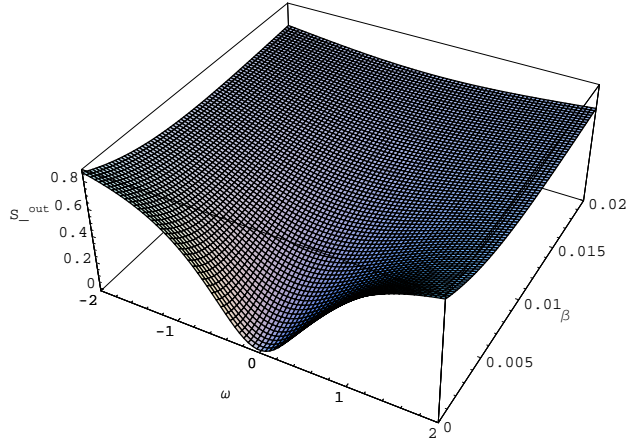}
\caption{A plot of the squeezing spectrum [Eq. \eqref{72b}] versus
$\beta$ and $\omega$ for $\kappa=0.8$ and $A=100$.}
\end{figure}

\begin{figure}[htbp]
\centering
\includegraphics [width=8.3cm]{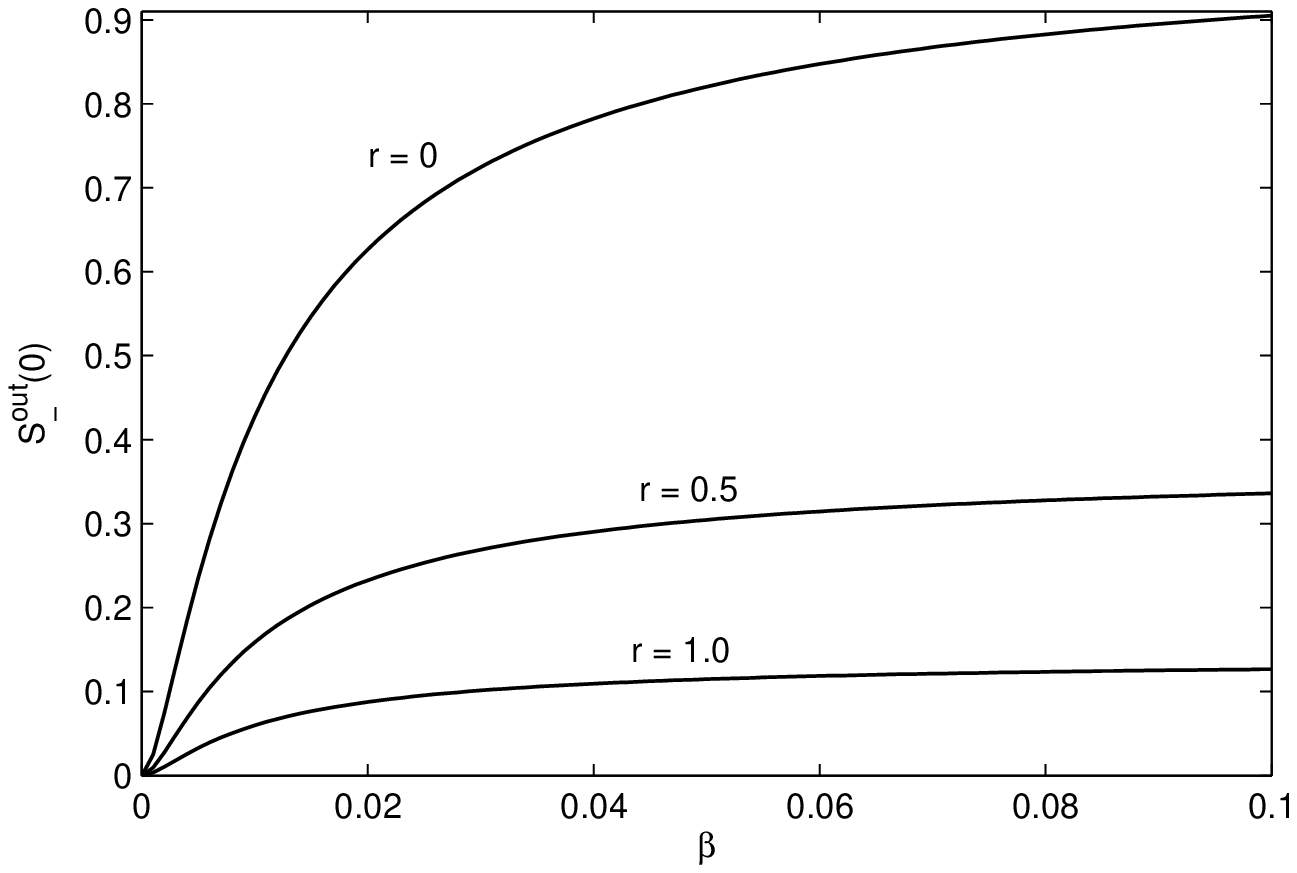}
\caption{Plots of the squeezing spectrum [Eq. \eqref{72b}] versus
$\beta$ for $\omega =0$, $\kappa=0.8$, $A=100$, and for different
values of the squeeze parameter $r$.}
\end{figure}

\begin{figure}[htbp]
\centering
\includegraphics [width=8.3cm]{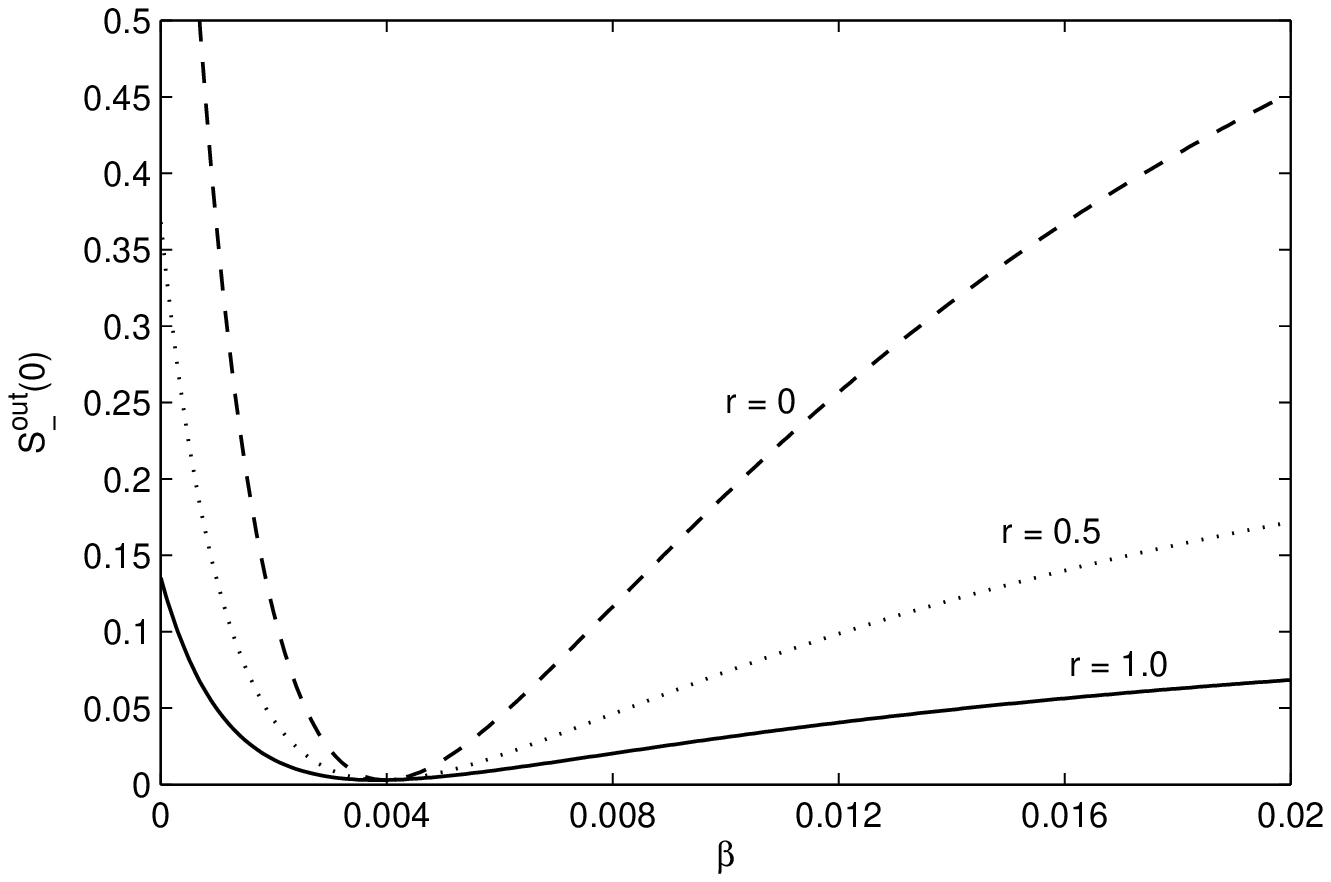}
\caption{Plots of the squeezing spectrum [\eqref{73b}] versus
$\beta$ for $\omega =0$, $\kappa=0.8$, $A=100$, and for different
values of the squeeze parameter $r$.}
\end{figure}

\begin{figure}[htbp]
\centering
\includegraphics [width=8.3cm]{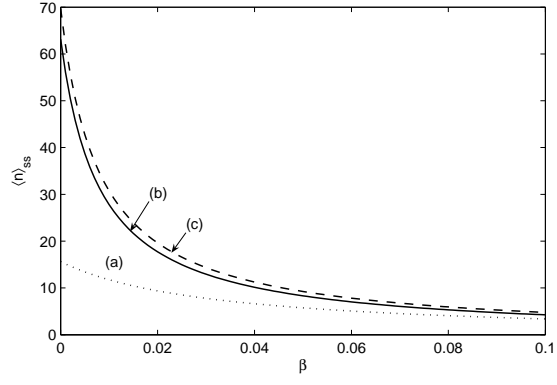}
\caption{Plots of the mean photon number [\eqref{77}] at steady
state versus $\beta$ for $\kappa=0.8$, $A=25$ and (a) for $r=0$
and $\varepsilon=\lambda\mu=0$ with $\mu\neq 0$, (b) for $r=0$ and
$\varepsilon=0.3$, (c) for $r=1.0$ and $\varepsilon=0.3$.}
\end{figure}

\begin{figure}[htbp]
\centering
\includegraphics [width=8.3cm]{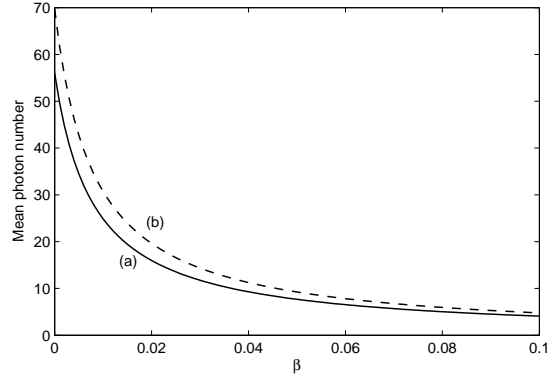}
\caption{(a) A plot of the mean photon number [\eqref{78}] at
steady state versus $\beta$ for $\kappa=0.8$, $A=25$, $r=1.0$, and
for $\varepsilon=0.3$. (b) A plot of the mean photon number
[\eqref{77}] at steady state versus $\beta$ for $\kappa=0.8$,
$A=25$, $r=1.0$, and for $\varepsilon=0.3$. }
\end{figure}

\begin{figure}[htbp]
\centering
\includegraphics[width=8.3cm]{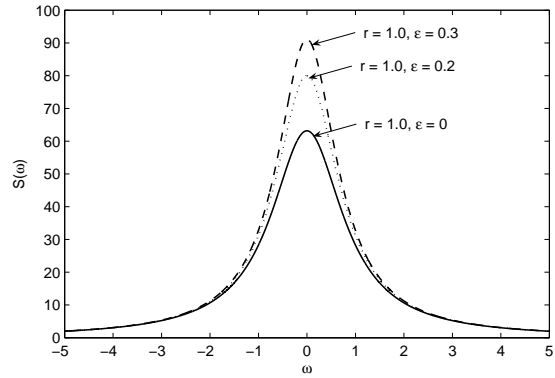}
\caption{Plots of the power spectrum
 [Eq.\eqref{86}] versus $\omega$ for A=100,
$\beta=0.01$, $r=1.0$, $\kappa=0.8$, and for different values of
$\varepsilon$.}
\end{figure}

\end{document}